# Quantitative Analysis of the Tumor/Metastasis System and its Optimal Therapeutic Control

Sébastien Benzekry, Dominique Barbolosi, Assia Benabdallah, Florence Hubert, Philip Hahnfeldt

*Center of Cancer Systems Biology, Steward Research and Specialty projects Corp., Tufts University School of Medicine, Boston, 02142, USA*

**Abstract**

A mathematical model for time development of metastases and their distribution in size and carrying capacity is presented. The model is used to theoretically investigate anti-cancer therapies such as surgery and chemical treatments (cytotoxic or anti-angiogenic), in monotherapy or in combination. Quantification of the effect of surgery on the size distribution of metastatic colonies is derived. For systemic therapies, emphasis is placed on the differences between the treatment of an isolated lesion and a population of metastases. Combination therapy is addressed, in particular the problem of the drugs administration sequence. Theoretical optimal schedules are derived that show the superiority of a metronomic administration scheme (defined as a continuous administration of a given amount of drug spread during the whole therapeutic cycle) on a classical Maximum Tolerated Dose scheme (where the dose is given as a few concentrated administrations at the beginning of the cycle), for the total metastatic burden in the organism.

**Introduction**

Metastases are the main cause of death in a cancer disease (1): cure rate goes from 90% all cancers combined when metastases are absent at diagnosis to 15% when they are present (2). They represent a major therapeutic challenge as prediction of their presence and development is limited by the resolution of imagery modalities only able to detect them when they reach a significant size (of the order of $10^8$ cells, i.e. approximately 100 mm$^3$). Clinical problems such as when to perform surgery of a primary lesion, if and how to administer adjuvant (i.e. after surgery) therapy or how to combine classical cytotoxic therapies with new biotargeted agents such as anti-angiogenic drugs are still open and would beneficiate from rationales based on theoretical studies of the metastatic process.

Mathematical modeling offers interesting tools that could give insights into a better understanding and control of these open clinical problems. However, most of the cancer modeling efforts are focused on tumoral development, often in a multi-scale framework aiming at compiling the large amount of knowledge about cancer biology. See (3) for a review of these models. Very few models are designed to quantify the metastatic process and development. In the 1970's, Liotta, Kleinerman and Saidel developed combined experimental and modeling approaches to study and quantify the various steps of metastasis creation. In (4), a deterministic model is proposed that describes tumor growth, interactions with the surrounding vasculature, dislodgement of tumor cells into the circulation, arrest in pulmonary vascular bed and formation of metastatic foci. The model is able to accurately describe experimental data of lung metastases spread by a fibrosarcoma and gives insights about perturbations of the system such as tumor massage, tumor resection, lung vessel damage, inhibition of vascularization or of tumor cell penetration of vessels. In (5) this approach is made stochastic in order to give interesting outputs such as the probability of metastases. More recently, Retsky and coworkers (6) developed another stochastic model in order to give a theoretical framework for study of the possible accelerating effect of surgery of a primary lesion on the growth of metastases. Simulations of the model are able to describe a bimodal pattern observed in a large data set of recurrence hazard in breast cancer, showing that the first peak of recurrence could be associated to a surgery-associated trauma. Another probabilistic model is proposed by Willis et al. (7) and used to study post-surgery dormancy in breast cancer.

Iwata, Kawasaki and Shigesada introduced in (8) a very interesting model for development of metastatic colonies, designed to describe the temporal development of the size distribution of a population of tumors. The size structure present in the model allows to describe for both visible metastases but also occult micro-metastases. Confrontation of the model to clinical data of metastatic growth in a patient with hepatocellular carcinoma showed good agreement and assessed the ability of the model to describe metastatic development. This model was further mathematically and numerically studied in (9,10), in particular with the intent to develop extensions of the model for cytotoxic therapy. Based on this approach and the model of Hahnfeldt, Panigraphy, Folkman and Hlatky (11) for tumoral growth under angiogenic control, a global model for metastatic development taking angiogenesis into account was developed (12–14). One of the main interest of this last model is to be able to simulate the action of anti-angiogenic therapy.

These modeling efforts form the ground basis of a theoretical study of the impact of scheduling of anticancer agents on the global dynamics of the disease. Indeed, not only the total dose of administered agent is of relevance for efficacy of a cancer treatment. For instance, metronomic chemotherapy (15–17) that gives cytotoxic agents at low dose but more continuously appears as a potential competitor of the more classical, maximum tolerated dose (MTD) administration protocols. In this context, scheduling of the drug is a critical player (18) that should be rationally optimized. Anti-angiogenic therapy (19) is also facing the same challenges. Impact of the administration scheme on the tumor growth has been evidenced in monotherapy (20,21) as well as in combination with chemotherapy (22), this last situation being the most common in the clinic.

Optimal control theory applied to cancer treatment has been developed since the pioneering work of Swan in the late 70's (23,24), first intended to optimize chemotherapy delivery under toxicity constraints. Various studies on chemotherapy were further conducted, for instance in the Model 1 project (25–29) that drove a phase I study for safe densification of a chemotherapeutic treatment in breast cancer. This model is focused on hematotoxicities due to the aggressive cytotoxic regimen. Using the Hahnfeldt model for AA treatment,

Ledzewicz and Schättler (30–32) studied optimal delivery of anti-vascular agents, revealing a singular structure in the optimal control. Scheduling implications of the Hahnfeldt model for AA monotherapy were also investigated by d'Onofrio and Gandolfi (33,34). Combination therapy involving AA agents was the subject of other works, either with radiotherapy (35) or chemotherapy (36). However, these models only deal with the primary tumor growth and don't take into account for the metastatic state of the disease.

**Modeling of the metastatic development**

We developed a mathematical model for description of the metastatic development integrating three major processes of the disease progression: proliferation, angiogenesis and metastatic spreading. The global philosophy is to place ourselves at the organism scale and to consider a growing population of secondary tumors (metastases). This approach was first initiated by Iwata et al. (8) that proposed a model for development of metastatic colonies structured by size, based on a gompertzian tumor growth rate. This model was further studied in (9,10) in order to establish well-posedness and efficient numerical methods for simulation of the model. This model was shown to accurately describe individual clinical data of hepatic metastases temporal evolution (8) and could yield an interesting prognostic tool to predict recurrence of the disease for breast cancer patients with unifocal lesion at diagnosis (37). Thanks to the size structure, effects of a systemic cytotoxic therapy can be added to the model and *in silico* simulations could help to determine the number of chemotherapeutic cycles to perform in order to avoid recurrence of the disease in a clinical setting, depending on the patient's cancer specific metastatic potential. However, this setting did not take angiogenesis into account and hence could not allow for simulation of anti-angiogenic therapy. The model that we present now, which was first developed in (14,38) merges the original model of (8) and the tumoral growth model under angiogenic control of Hahnfeldt et al. (11) in order to obtain a more general model for metastatic development taking angiogenesis into account.

We consider that tumors are biological entities with two phenotypical traits: volume (denoted by $V$, also referred as their size, expressed in mm³) and carrying capacity (denoted by $K$, expressed also in mm³). Primary tumor volume and carrying capacity are respectively denoted by $V_p$ and $K_p$. We assume that tumors have size and carrying capacity bigger than the size of one cell $V_0$ and smaller than a maximal reachable size $V_{max}$. Hence the physiological domain where metastases live is the square $\Omega = [V_0, V_{max}] \times [V_0, V_{max}]$, whose boundary is denoted by $\partial\Omega$ with external normal vector $\nu(\sigma)$. The metastatic population is mathematically represented by a density function $\rho(t,V,K) \in L^1(\Omega)$, $\forall t \in [0,T]$, where $T$ is the end time. This means that metastases live in a continuum of sizes and carrying capacities and that the number of tumors between sizes $V_1$ and $V_2$ and carrying capacities between $K_1$ and $K_2$ at time $t$ is given by

$\int_{V_1}^{V_2} \int_{K_1}^{K_2} \rho(t,V,K) dV dK$. We assume that each tumor grows with the same growth rate denoted by $G(V,K)$ without therapy and by $\bar{G}(V,K)$ when growth is perturbed by the action of a treatment, that spreading of new metastases is governed by an emission rate $\beta(V)$ and that distribution of metastases at birth is given by a function $N(\sigma)$ for $\sigma \in \partial\Omega$. Precise expressions of these functions will be described in the following. We also assume that there is no extinction: once created, a metastasis cannot exit the domain.

Overall, the model is a linear partial differential equation of renewal type, i.e. a transport equation endowed with a nonlocal boundary condition and an initial condition.

$$(1.1) \begin{cases} \partial_t \rho + \text{div}(\rho \bar{G}) = 0 & ]0,T[\times\Omega \\ -\bar{G}(t,\sigma) \cdot v(\sigma) \rho(t,\sigma) = N(\sigma) \left\{ \int_\Omega \beta(V) \rho(t,V,K) dV dK + \beta(V_p(t)) \right\} & ]0,T[\times\partial\Omega \\ \rho(0,V,K) = \rho^0(V,K) & \Omega \end{cases}$$

As a general modeling principle, we want to keep the number of parameters as low as possible.

*Tumor growth model*
For parsimony reasons, the growth rate of the tumors is supposed to be the same for the primary and for all the secondary tumors. We use the Hahnfeldt model (11) where sigmoidal tumor growth emerges from the interplay between a tumor and its carrying capacity, assumed to be representative of the vascular support provided to the neoplasm. The tumor compartment dynamics is gompertzian with a dynamical carrying capacity. The carrying capacity dynamics results from the balance between pro- and anti-tumor-induced angiogenic signaling. A study of diffusion-consumption equations for the concentration of angiogenic inhibitors and stimulators led Hahnfeldt et al. to the following expression

$$G(V,K) = \begin{pmatrix} aV \ln\left(\dfrac{K}{V}\right) \\ bV - dV^{2/3} K \end{pmatrix}$$

with $a$ being a parameter controlling the cancer cells proliferation kinetics, $b$ a parameter for production and effect of angiogenic stimulators and $d$ a parameter for production and effect of angiogenesis inhibitors. The main assumption underlying their model is that clearance rate of inhibitors (such as endostatin, angiostatin, thrombospondin-1,...) is much smaller than the clearance rate of stimulators (such as vascular endothelial growth factor, basic fibroblast growth factor,...). Analysis of the diffusion dynamics then implies that the concentration of inhibitors should be proportional to the surface of the tumor, hence explaining the 2/3 power in the inhibition term, while concentration of stimulators should be independent of tumor volume.

One of the main feature of this model is that, by taking angiogenesis into account *via* the interplay between the tumor and its vascular support, it is an appropriate framework to model the effect of an anti-angiogenic (AA) treatment, by an action on the carrying capacity. Following (11) we assume a linear killing term for the effect of an AA drug. Cytotoxic (CT) therapy is modeled using the log-kill assumption (39), which means that a cytotoxic drug kills a constant fraction of the cancer cells population. However, to satisfy the "no extinction" assumption that we made, we slightly modify these linear terms to ensure non-negativity of the entering flux $-\bar{G}(t,\sigma)\cdot v(\sigma)$ for all time. Denoting by $C(t)$ the concentration of CT agent, $f$ its efficacy parameter, by $A(t)$ the concentration of AA agent and $e$ its efficacy parameter, the expression of the growth rate under action of a combined treatment is given by

$$\bar{G}(V,K) = \begin{pmatrix} aV\ln\left(\dfrac{K}{V}\right) - fC(t)(V-V_0) \\ bV - dV^{2/3}K - eA(t)(K-K_0) \end{pmatrix}$$

Expressing the conservation of number of metastases when they are growing in size yields to the first equation of (1.1).

*Metastatic emission*
In the literature, there is no clear consensus about metastases being themselves able to metastasize or not. Here, we argue that there is no reason that cancer cells who acquired the ability to metastasize would loose it when establishing in a new site. Moreover, since metastasis is a long process before being detectable (in particular because tumors could remain dormant during possibly large time periods), the absence of clear proof in favor of metastases from metastases could be due to the short duration of the experiments compared to the time required for a secondary generation of tumors to reach a visible size. Here we are interested in long time behaviors and, although metastases from metastases could be neglected in first approximation, we think this second order term is relevant in our setting and chose to include it in our modeling, following clinical evidences of second-generation metastases (40).

Creation of new metastases is represented by the entering flux in the boundary condition of (1.1). It is the sum of two terms. Considering an emission rate $\beta(V)$ for a tumor with size $V$, the term $\beta(V_p(t))$ is the number of metastases emitted by the primary tumor, whereas the integral term $\int_\Omega \beta(V)\rho(t,V,K)dVdK$ is the number of metastases created by the metastases themselves.

The global metastatic process is complex and still not yet fully understood. Multiple steps are involved that include: detachment from the tumor, intravasation in blood or lymphatic vessels, survival in the circulation, escape from immune surveillance, extravasation, survival and settling in a new environment (41). For simplicity and parsimony, we don't give a detailed modeling of all these steps but rather consider, as done by Iwata et al. (8), a global emission rate that considers only the cells that succeeded in all the steps.

We take the same size-dependent expression as (8), and the amount of metastases emitted by a tumor with volume $V$ per unit of time is given by

$$\beta(V) = mV^\alpha$$

It depends on two parameters: the metastatic aggressiveness $m$ and $\alpha$ that is the third of the fractal dimension of the vasculature. The latter is assumed to be an intrinsic feature of the cancer and common to the development of all the tumors. It reflects that vasculature can be only superficial ($\alpha = 2/3$) or fully penetrating the tumor ($\alpha = 1$) or even having any fractal dimension between 2 and 3.

Metastases are assumed to be born with the size of one cell in view of the following remarks. Vascular holes by which a detaching cancer cell has to escape from the tumor have diameter of the order of 100 nanometers while the diameter of a cell is of the order of a micrometer. If a cell detaches from the tumor, it means that the cadherin (transmembrane proteins responsible for cell-cell adhesion) level falls down. Thus it seems unlikely that the cell that lost cadherins would still keep some to form a cluster. Even in the assumption of the detachment of a cluster of cells, it would be composed of at most a dozen of cells and the hypothesis of size 1 cell for the neo-metastasis would stay in a convenient approximation. All the metastases are assumed to be born with the same carrying capacity $K_0$. Indeed, data on the distribution of carrying capacity for metastatic newborns is not available in the literature and it was shown (42) that this assumptions represents the asymptotic limit of a concentrating distribution of metastases at birth $N(\sigma)$. Hence we take

$$N(\sigma) = \delta_{\sigma=(V_0, K_0)}$$

where $\delta$ stands for the Dirac mass.

**Simulations of metastatic development with and without systemic therapy**

*Parameter values*
Growth parameters of the Hahnfeldt model for Lewis Lung Carcinoma (LLC) implanted subcutaneously in C57BL/6 mice were estimated in (11) by fitting the model to average growth curves, using a Monte Carlo algorithm. We use these parameters for the growth of both the primary and the metastases. Metastatic parameters were arbitrarily fixed in order to give relevant biological values of the number and total burden of metastases. Parameter values are reported in the Table 1

| Parameter | Value | Unit | Meaning | Origin |
|---|---|---|---|---|
| a | 0.192 | day$^{-1}$ | Proliferation kinetics | (11) |
| b | 5.85 | day$^{-1}$ | Angiogenic stimulation | (11) |
| d | 0.00873 | day$^{-1}$mm$^{-2}$ | Angiogenic inhibition | (11) |
| m | 0.001 | day$^{-1}$ mm$^{-3\alpha}$ | Colonization rate | AF |
| α | 2/3 | | Fractal dimension of vascularization | AF |
| $K_0$ | 1 | mm$^3$ | Initial carrying capacity | AF |

Table 1: Parameter values. AF = Arbitrarily Fixed

Simulations of the model (1.1) were performed using to the numerical method developed in (12). It is a Lagrangian scheme based on the straightening of the characteristics. Euler or order 4 Runge-Kutta schemes were used for discretization of the characteristics and to solve ordinary differential equations when required.

*Cancer history and surgery*

We used the model to simulate the development of the metastatic population with the mice parameters described above, starting from the first cancer cell at time 0, and during 60 days (corresponding to approximately 17 years in human). The results are shown in Figure 1. Since the LLC is a fast growing tumor, the total metastatic burden is not very important compared to the volume of the primary tumor (Figure 1.A). At the end of the simulation, the metastases represent 1433 mm$^3$. An interesting feature of the model is the size structure that allows us to compute the number of visible metastases, i.e. the tumors with size bigger than a visibility threshold here taken to be 1 mm$^3$. This number can then be quantitatively compared to the total number of metastases that includes the occult micro-metastases (Figure 1.B). With the parameters that we used there is approximately three times more metastases in total than the mere visible ones. Our model allows a precise description of the size distribution of metastases at any given time. Figure 1.C shows this distribution at the end time. Knowledge of this distribution in a clinical setting could be of precious help in determining a precise diagnosis of a patient's metastatic state. However, this would of course require identification of its tumor growth and metastatic emission parameters, a process that is still under investigation.

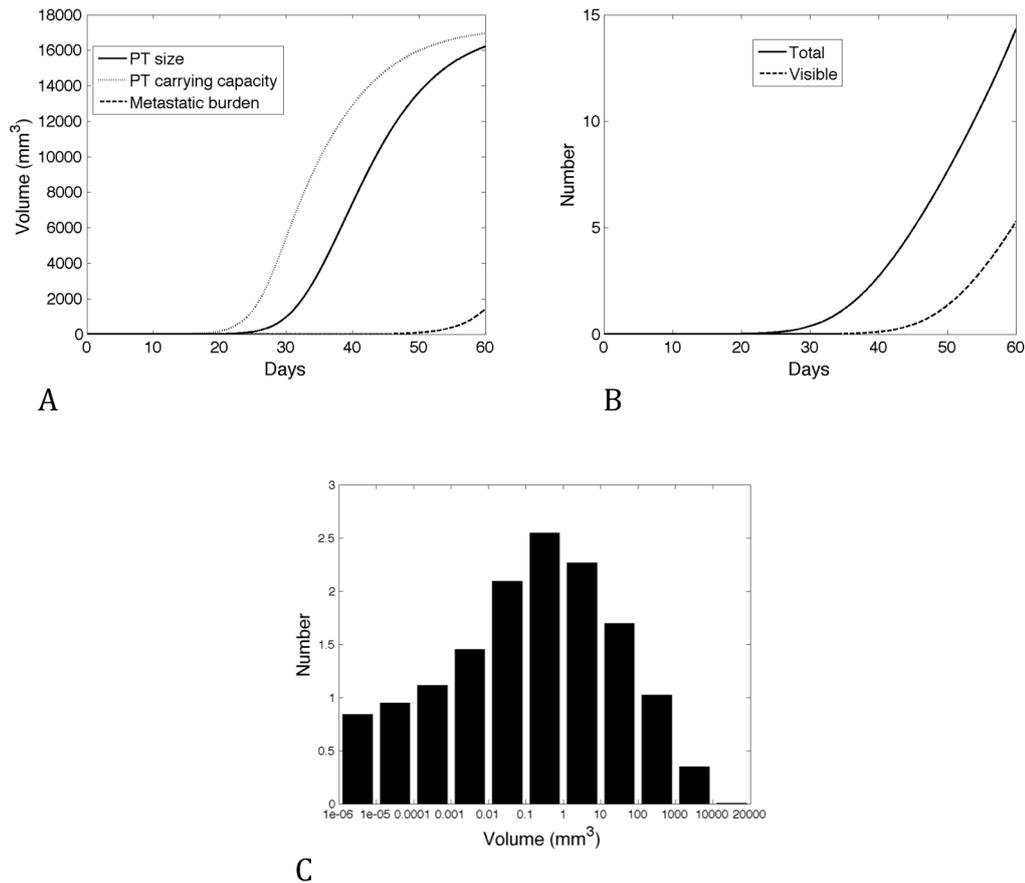

**Figure 1: Simulation of the cancer history from the first cancer cell. A. Primary tumor (PT) size and carrying capacity and metastatic burden B. Total and visible numbers of metastases. C. Size distribution of metastases at the end of the simulation**

In Figure 2 we present some simulations were *in silico* surgery of the primary lesion was performed when it reached 1500 mm$^3$ (31.6 days after the first cancer cell). As expected, resection stops the spreading process and results in a significant reduction of the number of metastases when compared to the situation without resection (Figure 2.A). Interestingly, our model reveals that this stop in the spreading would have almost no impact in the future development of the metastatic burden and would result in almost the same total secondary volume 60 days after cancer initiation, whether or not surgery is performed. Indeed, most of the metastatic burden comes from metastases that were created before the primary tumor reached 1500 mm$^3$ and continued growing after the resection. This result emphasizes the importance of adjuvant chemotherapy after surgery in order to treat the burden of invisible metastases.

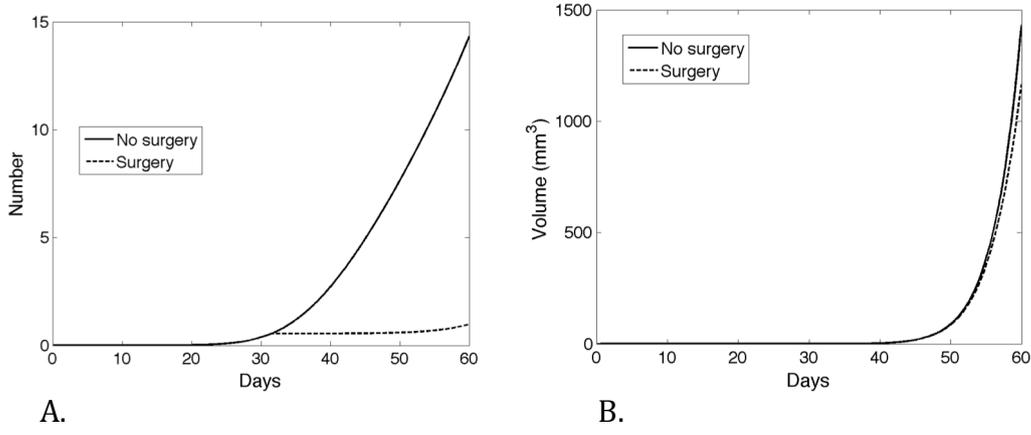

A.  B.

Figure 2: Simulation of surgery of the primary tumor when it reached 1500 mm³ and comparison with the case without surgery. A. Number of metastases. B. Metastatic burden

*Therapy*

By taking interactions of each tumor with its surrounding vasculature into account, the model is able to simulate the effect of AA drugs, both on the primary tumor and on the metastases. In (11) the effect of three AA agents were modeled *via* a linear first order pharmacokinetic elimination rate denoted $clr$ and an efficacy parameter $e$. The agent concentration is given by

$$A(t) = D \sum_{i=1}^{N} \exp(-clr(t-t_i)) \mathbf{1}_{t \geq t_i}$$

where $D$ is the administered dose, the $t_i$'s are the administration times and $\mathbf{1}_{t \geq t_i}$ is a Heaviside function having value one if and only if $t \geq t_i$. Efficacy and clearance parameters for three different agents, endostatin, angiostatin and TNP-470 were estimated in (11) from mice data and are reported in Table 2 for completeness.

| Agent | $clr$ (day⁻¹) | $e$ (day⁻¹conc⁻¹) | Protocol |
|---|---|---|---|
| Endostatin | 1.7 | 0.66 | 20mg/day |
| Angiostatin | 0.38 | 0.15 | 20mg/day |
| TNP-470 | 1.3 | 10.1 | 30mg/q.o.d |

Table 2: AA treatment parameters for three agents from (11)

Simulation results on the primary tumor growth and the number of metastases are shown in Figure 3. All three agents have different impacts on primary tumor growth and on metastatic spreading. Interestingly, the AA drug showing the best reduction of primary tumor growth when considering the size at the end of the simulation, namely angiostatin, is not the one provoking the best reduction in the number of metastases, which is endostatin. These two agents are given with the same scheduling but endostatin, having a more than four-fold larger efficacy parameter, implies better reduction of the primary tumor size during the treatment. This results in less emission of new metastases and thus less of them at the end of the simulation. However, due to its more than four-fold larger clearance rate compared to angiostatin, elimination is faster and regrowth after treatment cessation is more important.

This result suggests that different drug properties could result in different dynamical effects on the primary tumor and the metastases. The best drug could be different whether concern is focused on primary tumor reduction or control of the number of metastases.

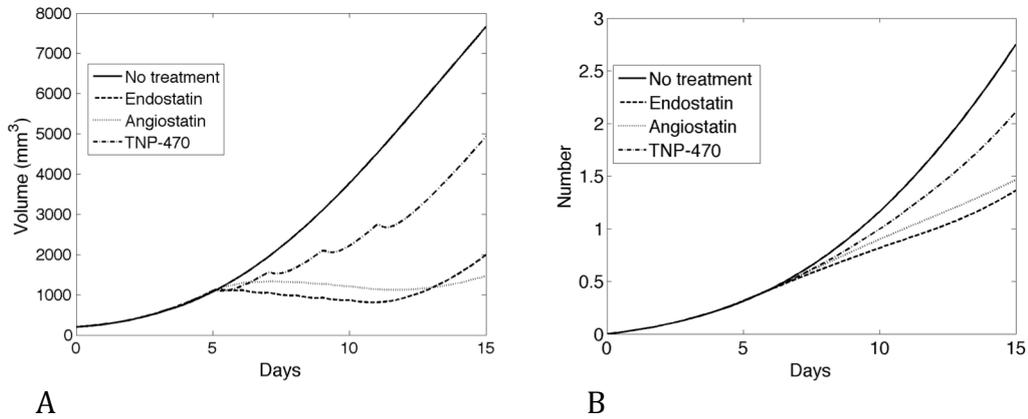

Figure 3: Effect of three anti-angiogenic treatments: endostatin 20mg/day, angiostatin 20mg/day, TNP-470 30mg/q.o.d. A. Primary tumor growth. B. Number of metastases

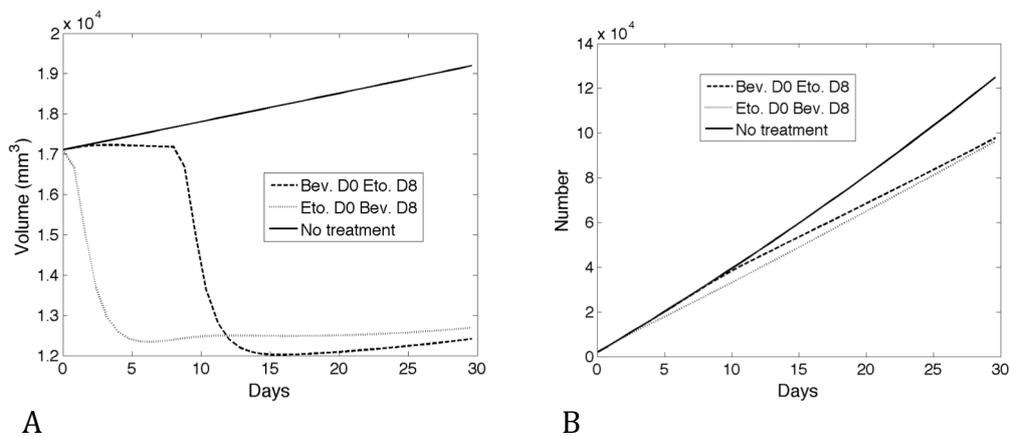

Figure 4: Combination therapy CT (etoposide) + AA (bevacizumab). Comparison of the order of administration: AA first on day 0 then CT on day 8 versus CT first on day 0 and then AA on day 8. Growth parameters taken to represent human tumor growth: a=0.0042, b=1, d=5.73x10$^{-4}$. Metastatic parameters: m=1, α=2/3, K$_0$=2630. Pharmacokinetic models and parameters for the drugs are from the literature. A. Primary tumor volume. B. Number of metastases.

We investigated then the combination of an AA drug and a CT one. To do so, we placed ourselves in a more clinical setting and considered growth parameters corresponding to the reported growth of an hepatocellular carcinoma (8) as well as more sophisticated pharmacokinetics models for the drugs. We focused on Etoposide for the cytotoxic agent, which is used in a wide variety of cancers (lung, testicle, lymphoma, leukemia,…) and Bevacizumab as our AA drug (monoclonal antibody targeting vascular endothelial growth factor, mainly used in colorectal and breast cancers). A few recent clinical trials have been evaluating this combination in lung cancers and glioblastoma, with mixed results (43–46). The pharmacokinetic models and parameters were taken from the literature: reference (25) for Etoposide and reference (47) for Bevacizumab.

Efficacy parameters $e$ and $f$ as well as metastatic emission parameters $m$ and $\alpha$ were arbitrarily fixed.

An important question in combining these two drugs is the order of administration: should the CT be given first and then the AA or reverse? Figure Figure 4 gives insight on this problem. We performed *in silico* simulations of both situations and compare the effects on the primary tumor development as well as on the number of metastases. On the primary tumor (Figure 4.A), our results suggest that it would be better to give first the AA drug, which has a stabilizing action and then the CT agent. Interestingly, this is in conjunction with the normalization theory (45) that proposes a pruning effect of AA drugs, which could potentiate the delivery of the chemotherapy. However, this feature is not explicitly included in the model (see (48,49) for modeling approaches of this phenomenon). On the other hand, effect on the metastatic spreading and total number of them is qualitatively more important when CT drug is administered first, which is the opposite strategy than found for the primary tumor.

These results suggest that order of administration of two drugs in a CT-AA combination setting could yield to different situations on the primary tumor and on the metastases.

**Scheduling optimization**

*Formulation of an optimization problem*

As seen in the previous examples, the same drug or combination of drugs can show different efficacy on primary tumor reduction and metastatic limitation. In order to perform a rational theoretical study of these differences, we define now an optimization problem for the metastases.

We focus here on the AA monotherapy situation and consider that a total amount $A_{max}$ has to be used at a constant rate during some administration duration $\tau$ followed by a rest period from $\tau$ to an arbitrary end time $T$, i.e. $A(t;\tau) = \dfrac{A_{max}}{\tau}\mathbf{1}_{t \leq \tau}$, in the same way as (31). We are then looking at optimizing the treatment duration $\tau$ - that we will also refer to as the scheduling strategy - regarding to various objectives. The two extreme strategies correspond to the clinical situations of a Maximum Tolerated Dose (MTD) administration scheme, where most of the drug is given at a strong dose during a short time, followed by a recovery period until start of the next cycle, and the metronomic administration scheme, where dose is spread out on the whole cycle, in a low-dose/large duration fashion. See Figure 5 for illustration. We will theoretically compare these two extreme scheduling strategies as well as all possible intermediate situations.

For the primary tumor dynamics, we use the Hahnfeldt model and define two objectives subject to be minimized under the action of one drug in monotherapy, or two drugs in combination. These two criteria are the size of the tumor at a fixed end time $T$, denoted by $J_T$, that takes into account possible regrowth of the tumor after cessation of the treatment, and the minimal size reached during the simulation interval [0,T], denoted by $J_m$. We denote by $V_p(t;\tau)$ the primary tumor volume at time $t$ when therapy is administered with rate $u$. Mathematical definitions of the two tumoral objectives are given by

$$J(\tau) = V_p(T;\tau) \quad \text{and} \quad J_m(\tau) = \min_{t \in [0,T]} V_p(t;\tau)$$

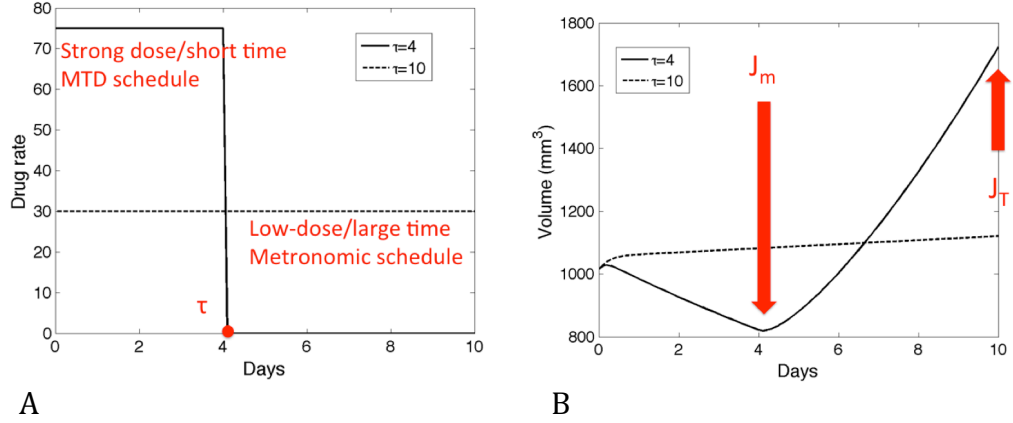

**Figure 5: The two extreme strategies for duration of the treatment: MTD or metronomic. A.** AA drug rate for the two extreme values of $t_A$. **B.** The two corresponding curves for primary tumor growth under AA treatment. The $t_A=4$ curve is used to illustrate the two objectives $J_m$ and $J_T$ used in the text for evaluation of treatment efficacy on the primary tumor.

For a given treatment strategy $\tau$, we define two additional objectives on the metastases: the total number of metastases $J$ and the total metastatic burden $J_M$, by

$$J(\tau) = \int_\Omega \rho(T,V,K,)dVdK, \quad \text{and} \quad J_M(\tau) = \int_\Omega V\rho(T,V,K)dVdK$$

For the simulations we used the following values of the parameters:
$$A_{max} = 300, \quad a = 0.084, \quad b = 5.85, \quad d = 0.00873, \quad m = 0.001, \quad \alpha = 2/3$$
for consistency with (31). The initial conditions for the primary tumor and the metastases were

$$V_{0,p} = 1015, \quad K_{0,p} = 6142, \quad V_0 = 10^{-6} (1 \text{ cell}), \quad K_0 = 625$$

*Simulations*
Although the total amount of drug is kept constant, changing the administration strategy has an important impact on all of the four objectives defined above. The effect of varying the scheduling strategy on the growth of the primary tumor is represented in Figure 6 where we observe different behaviors interpolating between the two extreme cases shown in Figure 5.B. The MTD strategy shows a sharp decrease of tumor size during the treatment period, but then a fast regrowth. On the opposite, the metronomic strategy has a more stabilizing effect that results in an overall end volume smaller than with the MTD strategy. It appears clearly on Figure 6 that the best strategy for the objective $J_T$ (tumor size at the end) is then the metronomic strategy whereas the best strategy for objective $J_m$ (minimal reached size during $[0,T]$) is the MTD one. These results are confirmed when plotting the objective values for different values of the scheduling strategy $\tau$ (see Figure 7.A). The result for $J_T$ was observed not to depend on parameter values neither initial conditions in a large range of simulations, whereas the result for $J_m$ did.

Results for the metastatic objectives are plotted in Figure 7.B. While the total metastatic burden $J_M$ exhibits the same qualitative behavior as the end tumor size $J_T$, the number of metastases $J$ suggests a nontrivial optimal scheduling administration $\tau^* = 6.5$ days. This is a strategy different from the MTD or the metronomic one. It confirms the previous results showing possible differences in the optimal treatment of the primary tumor and the metastases. Indeed, it is now numerically proven that the best strategy for minimizing metastatic emission is different from the best strategy for reduction of the primary tumor volume.

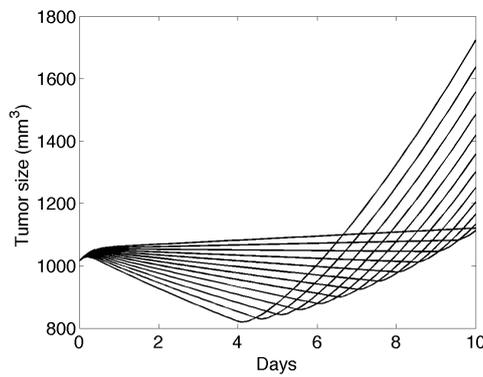

Figure 6: Primary tumor growth curves for various scheduling strategies τ

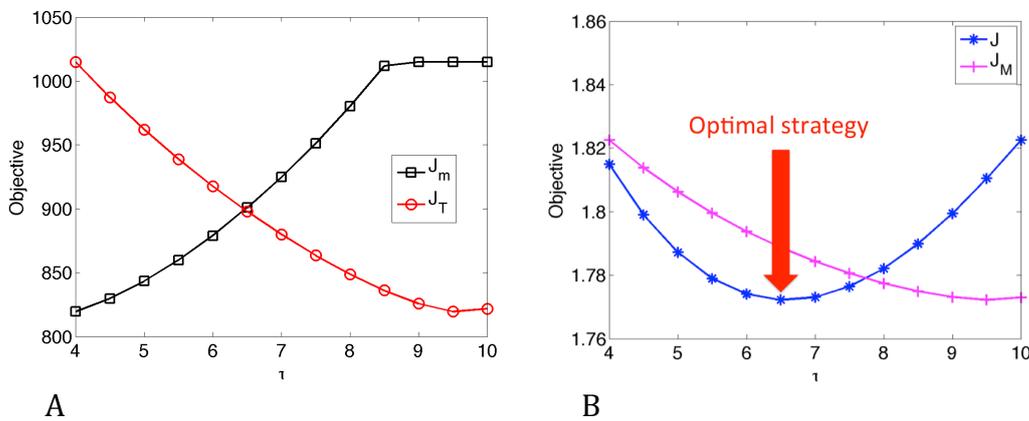

A                                      B

Figure 7: Various objectives against scheduling strategy. A. Primary tumor objectives. B. Metastatic objectives. y-axis scale only valid for the number of metastases.

*A model for low dose anti-angiogenic chemotherapy*

So far the model that we used for tumor growth did not take development of resistances to the treatment into account, although it is a major limitation in the efficacy of chemotherapy, especially in the case of MTD schedules. The genetically instable population of cancer cells rapidly evolves to form resistant clones that are insensitive to cytotoxic treatment, allowing the tumor to escape therapeutic control. However, when administered continuously at low dose, cytotoxic agent are shown to have an anti-angiogenic action (18,50). For example low dose Vinblastine has been shown (15) to induce more important reduced proliferation of endothelial cells than for cancer cells. This switch in the target to the more genetically stable endothelium was hence proposed to avoid the development of resistances and provoke better long term anti-tumoral effect.

To test these hypotheses on the dynamical behavior of a tumor in interaction with its surrounding vasculature, we integrated the resistance phenomenon in the Hahnfeldt model as well as a pharmacokinetic (PK)/pharmacodynamics (PD) model first intended for hematological toxicities (51). The main assumptions are: a) the CT drug has an anti-angiogenic effect by killing proliferative endothelial cells, b) cancer cells develop resistances whereas endothelial cells don't and c) the killing action of the drug is stronger on the endothelial compartment than on the tumoral one. The tumor growth rate is given by

$$\bar{G}(t,V,K) = \begin{pmatrix} aV\ln\left(\dfrac{K}{V}\right) - C_1(t)(V - V_{min})^+ \\ bV - dV^{2/3}K - C_2(t)(K - K_{min})^+ \end{pmatrix}$$

where $C_1$ and $C_2$ are the exposures of the CT drug respectively on the tumor cells and on the endothelial cells and $V_{min}$ and $K_{min}$ are minimal values of the tumor volume and carrying capacity for the drug to be active. Exposures are defined from the output $C(t)$ of the PK/PD model, whose equations and parameters can be found in (13) (the drug we consider here is Docetaxel, used in the treatment of breast cancer). They take into account assumptions b) and c) and their expressions are given by

$$C_1(t) = \alpha_1 e^{-R\int_0^t C(s)ds} C(t), \quad C_2(t) = \alpha_2 C(t).$$

with $R$ being the resistance parameter (each cancer cell has a probability $RC(t)$ per unit of time of becoming resistant when exposed to an amount $C(t)$ of drug) and $\alpha_2 > \alpha_1$.

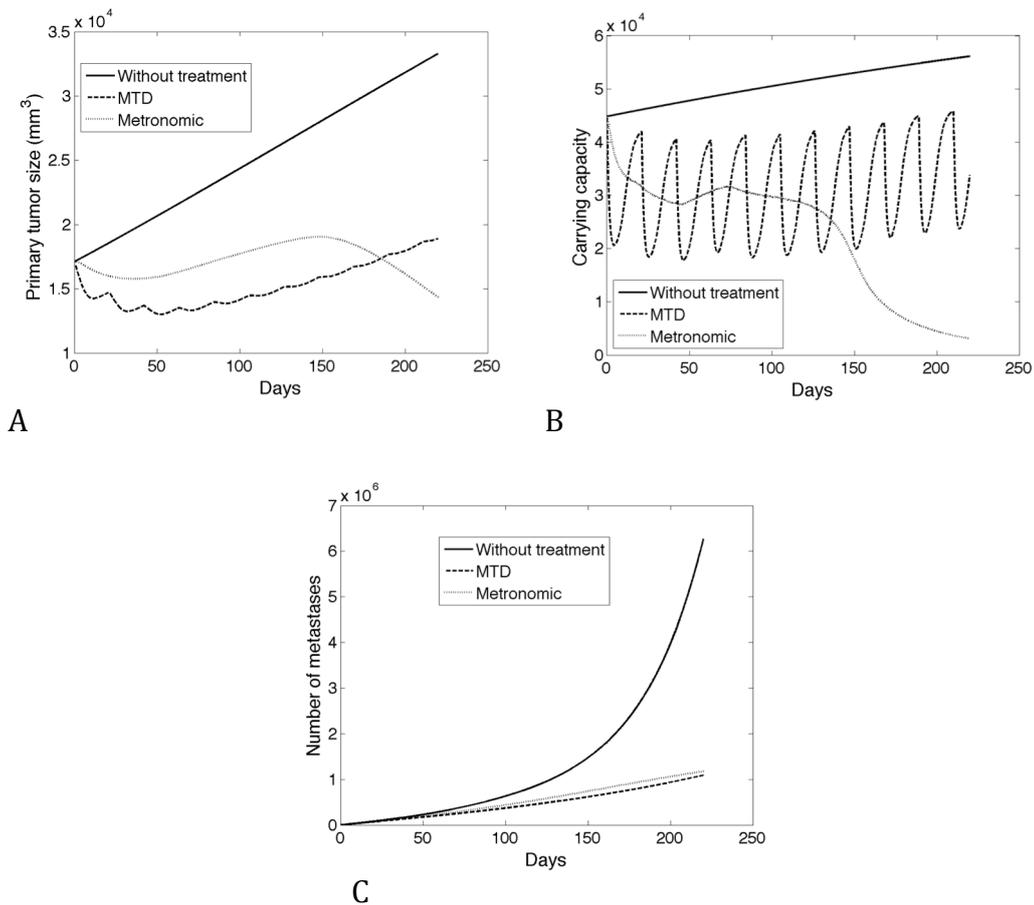

**Figure 8: Simulations of a maximum tolerated dose (MTD) and a metronomic protocol for chemotherapy. A Effect on tumor volume. B Effect on the carrying capacity. C Effect on the total number of metastases**

We used the model to test *in silico* the dynamical differences of two temporal administration schemes. The MTD protocol is composed of 21 days cycles with administration of 100 mg of Docetaxel on day 0 (51). The metronomic protocol gives 10 mg/day every day, without resting periods. Notice that the metronomic schedule has a higher total dose over one cycle (210 mg versus 100 mg for the MTD) but this is consistent with clinical practice, such as in (52). Simulations comparing the effect of these two protocols are shown in Figure 8. During the first cycles of the therapy, MTD scheduling exhibits a better tumor reduction (Figure 8.A), being able to significantly reduce the size of the tumor. Unfortunately, after the third cycle, regrowth of the tumor is observed, due to the development of resistances. On the other hand, the metronomic protocol shows very mixed results during the same period with even tumoral progression. However, on the long term (after about 150 days), the metronomic schedule shows its superiority, provoking a more pronounced and sustained tumor decrease. The differences between the two situations can be understood by looking at the dynamics of the carrying capacity in Figure 8.B (which represents the vasculature in the model). Both schedules have an impact on the endothelium however, in the MTD case, rapid regrowth of vascular support is observed after cessation of the treatment, for each cycle. Indeed, presence of the tumor stimulates this vascular recovery. In the metronomic situation, although the vascular injury has a lower amplitude, the effect is more sustained and induces a continuous decrease of the vascular support that eventually suffocates the tumor. Moreover, this effect does not induce resistances and continues for large times. Effect on the metastatic emission (Figure 8.C) is comparable for both protocols, although the MTD schedule is qualitatively better, resulting in slightly less total number of metastases than the metronomic.

These results support the hypothesis of a better effect of metronomic schedules over MTD due to differences in the dynamical recovery of the vasculature.

**Conclusion**

A mathematical model for metastatic growth was developed that describes the development of a population of metastases at the organism scale. Fundamental aspects of a cancer disease are in the model: cellular proliferation, vascular development and metastatic spreading, as well as three of the four major anti-cancer therapeutic tools: surgery, chemotherapy and anti-angiogenic therapy (but not radiotherapy). The model appears as a theoretical and numerical tool describing global temporal development of a cancer disease and its control. It gives an interesting framework for the study of scheduling strategies for chemical treatments, in particular for the differences between treatment of an isolated primary lesion and a generalized metastatic disease.

**References**

1.  Fidler IJ, Paget S. The pathogenesis of cancer metastasis: the "seed and soil" hypothesis revisited. Nat. Rev. Cancer. 2003;3(6):453–8.

2.  Chiarella P, Bruzzo J, Meiss RP, Ruggiero RA. Concomitant tumor resistance. Cancer Lett. Elsevier Ireland Ltd; 2012 May 22;324(2):133–41.

3.  Araujo RP, McElwain DLS. A history of the study of solid tumour growth: the contribution of mathematical modelling. Bull. Math. Biol. 2004 Sep;66(5):1039–91.

4.  Saidel GM, Liotta LA, Kleinerman J. System dynamics of metastatic process from an implanted tumor. J. Theor. Biol. 1976;56:417–34.

5.  Liotta LA, Saidel GM, Kleinerman J. Stochastic model of metastases formation. Biometrics. 1976 Sep;32(3):535–50.

6.  Retsky MW, Demicheli R, Swartzendruber DE, Bame PD, Wardwell RH, Bonadonna G, et al. Computer simulation of a breast cancer metastasis model. Breast Cancer Res. Treat. 1997 Sep;45(2):193–202.

7. Willis L, Alarcón T, Elia G, Jones JL, Wright NA, Tomlinson IP, et al. Breast cancer dormancy can be maintained by small numbers of micrometastases. Cancer Res. 2010 Jun 1;70(11):4310–7.

8. Iwata K, Kawasaki K, N. S, Shigesada N. A dynamical model for the growth and size distribution of multiple metastatic tumors. J Theor Biol. 2000 Mar 21;203(2):177–86.

9. Barbolosi D, Benabdallah A, Hubert F, Verga F. Mathematical and numerical analysis for a model of growing metastatic tumors. Math. Biosci. 2009 Mar;218(1):1–14.

10. Devys A, Goudon T, Lafitte P. A model describing the growth and the size distribution of multiple metastatic tumors. Discrete and Continuous Dynamical Systems - Series B. 2009 Aug;12(4):731–67.

11. Hahnfeldt P, Panigraphy D, Folkman J, Hlatky L. Tumor development under angiogenic signaling: a dynamical theory of tumor growth, treatment, response and postvascular dormancy. Cancer Res. 1999 Oct 1;59(19):4770–5.

12. Benzekry S. Mathematical and numerical analysis of a model for anti-angiogenic therapy in metastatic cancers. ESAIM: Mathematical Modelling and Numerical Analysis. 2011 Oct 5;46(2):207–37.

13. Benzekry S, André N, Benabdallah A, Ciccolini J, Faivre C, Hubert F, et al. Modelling the impact of anticancer agents on metastatic spreading. Math Model Nat Phenom. 2012;7(1):306–36.

14. Benzekry S. Mathematical analysis of a two-dimensional population model of metastatic growth including angiogenesis. J Evol Equ. 2011 Dec 23;11(1):187–213.

15. Klement G, Baruchel S, Rak J, Man S, Clark K, Hicklin DJ, et al. Continuous low-dose therapy with vinblastine and VEGF receptor-2 antibody induces sustained tumor regression without overt toxicity. J. Clin. Invest. 2000 Apr;105(8):15–24.

16. Kerbel RS. Continuous low-dose anti-angiogenic/metronomic chemotherapy: from the research laboratory into the oncology clinic. Annals of Oncology. 2002 Jan 19;13(1):12–5.

17. Pasquier E, Kavallaris M, Andre N. Metronomic chemotherapy: new rationale for new directions. Nat Rev Clin Oncol. 2010;7:455–65.

18. Browder T, Butterfield CE, Kräling BM, Kra BM, Shi B, Marshall B, et al. Antiangiogenic scheduling of chemotherapy improves efficacy against experimental drug-resistant cancer. Cancer Res. 2000;60:1878–86.

19. Folkman J. Anti-angiogenesis: new concept for therapy of solid tumors. Annals of surgery. 1972 Mar;175(3):409–16.

20. Drixler TA, Rinkes IHMB, Ritchie ED, Gebbink MFBG, Voest EE, Borel Rinkes IH, et al. Continuous Administration of Angiostatin Inhibits Accelerated Growth of Colorectal Liver Metastases after Partial Hepatectomy. Cancer Res. 2000 Mar;60:1761–5.

21. Reynolds AR. Potential relevance of bell-shaped and U-shaped dose-responses for the therapeutic targeting of angiogenesis in cancer. Dose-Response. 2009 Jan;8(3):253–84.

22. Gasparini G, Longo R, Fanelli M, Teicher B a. Combination of Antiangiogenic Therapy With Other Anticancer Therapies: Results, Challenges, and Open Questions. Journal of Clinical Oncology. 2005 Feb 20;23(6)(6):1295–311.

23. Swan GW. Role of optimal control theory in cancer chemotherapy. Math. Biosc. 1990 Oct;101(2):237–84.

24. Swan GW, Vincent TL. Optimal control analysis in the chemotherapy of IgG multiple myeloma. Bull. Math. Biol. 1977;39(3):317–37.

25. Barbolosi D, Freyer G, Ciccolini J, Iliadis A. Optimisation de la posologie et des modalités d'administration des agents cytotoxiques à l'aide d'un modèle mathématique. Bulletin du Cancer. 2003;90(2):167–75.

26. Meille C, Gentet JC, Barbolosi D, Andre N, Doz F, Iliadis A, et al. New adaptive method for phase I trials in oncology. Clin. Pharmacol. Ther. 2008 Jun;83(6):873–81.

27. Barbolosi D, Iliadis A. Optimizing drug regimens in cancer chemotherapy: a simulation study using a PK--PD model. Comput. Biol. Med. 2001 May;31(3):157–72.

28. Iliadis A, Barbolosi D. Optimizing drug regimens in cancer chemotherapy by an efficacy-toxicity mathematical model. Comput. Biomed. Res. 2000;33:211–26.

29. Iliadis A, Barbolosi D. Dosage regimen calculations with optimal control theory. Int. J. Biomed. Comput. 1994;36:87–93.

30. Ledzewicz U, Schättler H. Antiangiogenic therapy in cancer treatment as an optimal control problem. SIAM J. Control Optim. 2007;46(3):1052–79.

31. Ledzewicz U, Marriott J, Maurer H, Schättler H. Realizable protocols for optimal administration of drugs in mathematical models for anti-angiogenic treatment. Mathematical medicine and biology : a journal of the IMA. 2010 Jun;27(2):157–79.

32. Ledzewicz U, Munden J, Schättler H. Scheduling of angiogenic inhibitors for Gompertzian and logistic tumor growth models. Discrete Contin. Dyn. Syst. Ser. B. 2009;12(2):415–38.

33. D'Onofrio a, Gandolfi A, Rocca A. The dynamics of tumour-vasculature interaction suggests low-dose, time-dense anti-angiogenic schedulings. Cell Prolif. 2009 Jun;42(3):317–29.

34. D'Onofrio A, Gandolfi A. Tumour eradication by antiangiogenic therapy: analysis and extensions of the model by Hahnfeldt et al. (1999). Math. Biosci. 2004 Oct;191(2):159–84.

35. Ergun A, Camphausen K, Wein LM. Optimal scheduling of radiotherapy and angiogenic inhibitors. Bull. Math. Biol. 2003 May;65(3):407–24.

36. D'Onofrio A, Ledzewicz U, Maurer H, Schättler H. On optimal delivery of combination therapy for tumors. Math. Biosci. Elsevier Inc.; 2009 Nov;222(1):13–26.

37. Barbolosi D, Verga F, Benabdallah A, Hubert F, Mercier C, Ciccolini J, et al. Modélisation du risque d'évolution métastatique chez les patients supposés avoir une maladie localisée. Oncologie. 2011;13(8):528–33.

38. Benzekry S. Mathematical and numerical analysis of a model for anti-angiogenic therapy in metastatic cancers. M2AN. 2012;46(2):207–37.

39. Skipper H. Kinetics of mammary tumor cell growth and implications for therapy. Cancer. 1971 Dec;28(6):1479–99.

40. August DA, Sugarbaker PH, Schneider PD. Lymphatic dissemination of hepatic metastases. Implications for the follow-up and treatment of patients with colorectal cancer. Cancer. 1985 Apr;55(7):1490–4.

41. Gupta GP, Massagué J. Cancer metastasis: Building a framework. Cell. 2006 Nov 17;127(4):679–95.

42. Benzekry S. Passing to the limit 2D–1D in a model for metastatic growth. J Biol Dynam. 2012 Jan;6(sup1):19–30.

43. Spigel DR, Townley PM, Waterhouse DM, Fang L, Adiguzel I, Huang JE, et al. Randomized Phase II Study of Bevacizumab in Combination With Chemotherapy in Previously Untreated Extensive-Stage Small-Cell Lung Cancer: Results From the SALUTE Trial. J. Clin. Oncol. 2011;29:2215–22.

44. Reardon DA, Desjardins A, Peters K, Gururangan S, Sampson J, Rich JN, et al. Phase II study of metronomic chemotherapy with bevacizumab for recurrent glioblastoma after progression on bevacizumab therapy. J. Neurooncol. 2011;103:371–9.

45. Jain RK. Normalizing tumor vasculature with anti-angiogenic therapy: A new paradigm for combination therapy. Nature Medicine. 2001 Sep;7(9):987–9.

46. Francesconi AB, Dupre S, Matos M, Martin D, Hughes BG, Wyld DK, et al. Carboplatin and etoposide combined with bevacizumab for the treatment of recurrent glioblastoma multiforme. Journal of clinical neuroscience : official journal of the Neurosurgical Society of Australasia. Elsevier Ltd; 2010 Aug;17(8):970–4.

47. Lu J-F, Bruno R, Eppler S, Novotny W, Lum B, Gaudreault J. Clinical pharmacokinetics of bevacizumab in patients with solid tumors. Cancer chemotherapy and pharmacology. 2008 Oct;62(5):779–86.

48. Benzekry S, Chapuisat G, Ciccolini J, Erlinger A, Hubert F. A new mathematical model for optimizing the combination between antiangiogenic and cytotoxic drugs in oncology. Comptes Rendus de l'Académie des Sciences - Mathématiques. Elsevier Masson SAS; 2012 Jan;350(1-2):23–8.

49. d'Onofrio A, Gandolfi A. Chemotherapy of vascularised tumours: role of vessel density and the effect of vascular.

50. Kerbel RS, Kamen BA. The anti-angiogenic basis of metronomic chemotherapy. Nature Reviews Cancer. 2004 Jun;4(6):423–36.

51. Meille C, Iliadis A, Barbolosi D, Frances N, Freyer G. An interface model for dosage adjustment connects hematotoxicity to pharmacokinetics. Journal of pharmacokinetics and pharmacodynamics. 2008 Dec;35(6):619–33.

52. Baruchel S, Diezi M, Hargrave D, Stempak D, Gammon J, Moghrabi A, et al. Safety and pharmacokinetics of temozolomide using a dose-escalation, metronomic schedule in recurrent paediatric brain tumours. Eur. J. Cancer. 2006 Sep;42(14):2335–42.